\documentclass[aps,prl,
   amsfonts,amssymb,amsmath,
   twocolumn,
   superscriptaddress,
   noshowpacs,
   preprintnumbers,
   amsmath,amssymb
]{revtex4-1}

\pdfoutput=1
\usepackage{color}
\usepackage{graphicx}

\begin{document}


\title{Fermi-Surface Reconstruction and Complex Phase Equilibria in CaFe$_{2}$As$_{2}$}

\author{K. Gofryk}
\affiliation{Oak Ridge National Laboratory, Oak Ridge, Tennessee 37831, USA}
\author{B.~Saparov}
\affiliation{Oak Ridge National Laboratory, Oak Ridge, Tennessee 37831, USA}
\author{T.~Durakiewicz}
\affiliation{Los Alamos National Laboratory, Los Alamos, New Mexico 87545, USA}
\author{A.~Chikina}
\affiliation{Institute of Solid State Physics, Dresden University of Technology, Zellescher Weg 16, D-01062 Dresden, Germany}
\author{S.~Danzenb\"{a}cher}
\affiliation{Institute of Solid State Physics, Dresden University of Technology, Zellescher Weg 16, D-01062 Dresden, Germany}
\author{D.~V.~Vyalikh}
\affiliation{Institute of Solid State Physics, Dresden University of Technology, Zellescher Weg 16, D-01062 Dresden, Germany}
\author{M.~J.~Graf}
\affiliation{Los Alamos National Laboratory, Los Alamos, New Mexico 87545, USA}
\author{A.~S.~Sefat}
\affiliation{Oak Ridge National Laboratory, Oak Ridge, Tennessee 37831, USA}


\begin{abstract}

Fermi-surface topology governs the relationship between magnetism and superconductivity in iron-based materials. Using low-temperature transport, angle-resolved photoemission, and x-ray diffraction we show unambiguous evidence of large Fermi surface reconstruction in CaFe$_{2}$As$_{2}$ at magnetic spin-density-wave and nonmagnetic collapsed-tetragonal ($cT$) transitions. For the $cT$ transition, the change in the Fermi surface topology has a different character with no contribution from the hole part of the Fermi surface. In addition, the results suggest that the pressure effect in CaFe$_{2}$As$_{2}$ is mainly leading to a rigid-band-like change of the valence electronic structure. We discuss these results and their implications for magnetism and superconductivity in this material.

\end{abstract}

\maketitle

\begin{figure}[b!]
\begin{centering}
\includegraphics[width=0.35\textwidth]{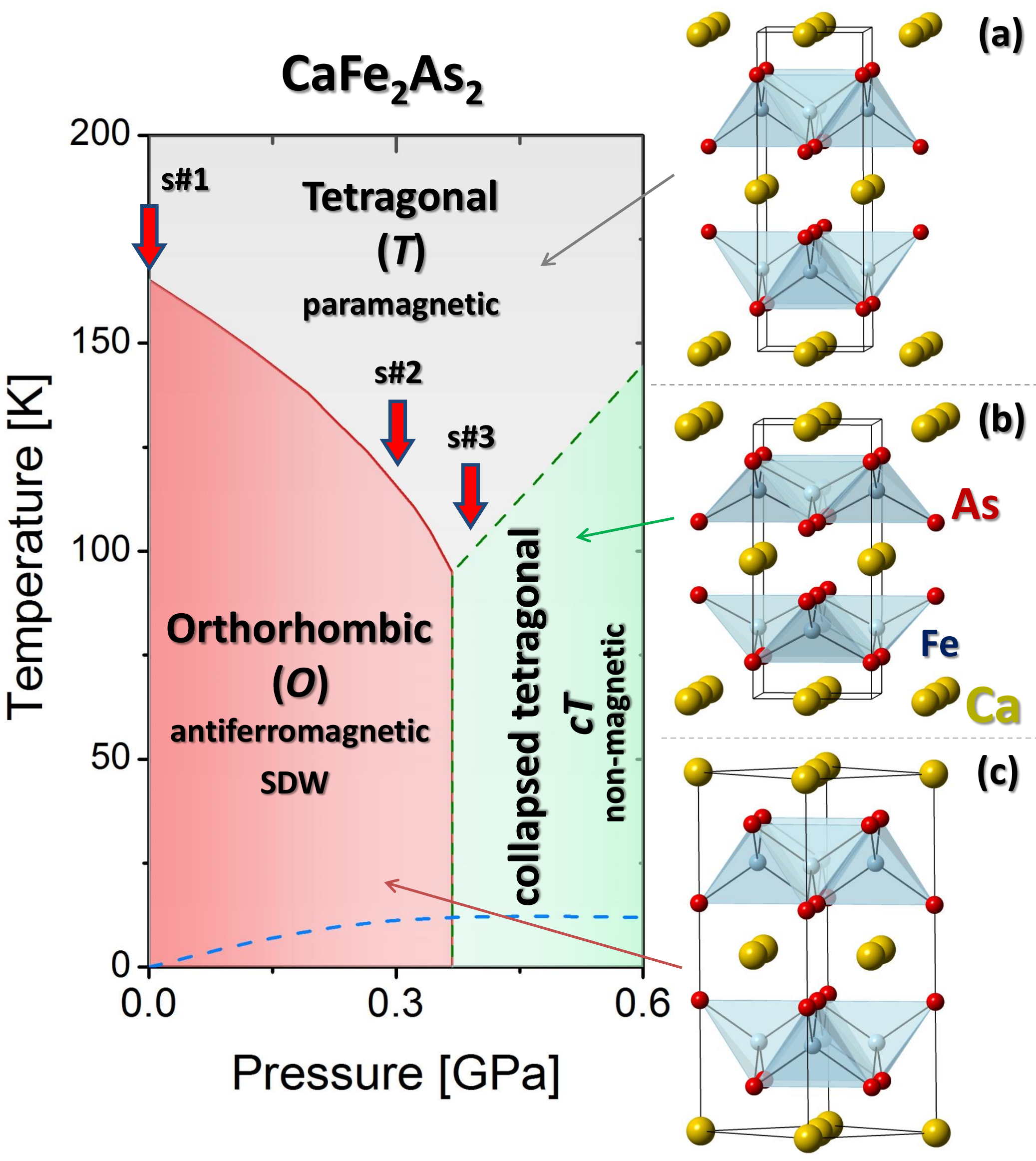}
\caption{(Color online) A putative phase diagram of CaFe$_{2}$As$_{2}$. The data are adapted from Refs.~\onlinecite{9,PC,PC2,hl}. The blue dashed line marks superconductivity in CaFe$_{2}$As$_{2}$ observed under non-hydrostatic pressure condition \cite{9}. On the right side, the crystal structures of (a) tetragonal (s.g. \emph{I4/mmm}), (b) collapsed tetragonal and (c) orthorhombic (s.g. \emph{Fmmm}) phases are presented. The structures are drawn with modified parameters in order to emphasize the differences between the three phases. The thick solid red arrows mark the transition temperatures of samples studied in this work (see text).}\label{1}
\end{centering}
\end{figure}

CaFe$_{2}$As$_{2}$ has a special position among the ThCr$_{2}$Si$_{2}$-type $A$Fe$_{2}$As$_{2}$ ($A$ = Ba, Sr, Ca) superconducting phases \cite{122a,122b,122c} due to complex relationship between magnetism, superconductivity and lattice instabilities (see Refs.~\onlinecite{gold,karel,mishra}). The way CaFe$_{2}$As$_{2}$ is typically grown under ambient pressure makes it crystallize in the tetragonal ($T$) crystal structure at room temperature, which then undergos a simultaneous spin-density-wave (SDW) and orthorhombic ($O$) transition at $\sim$170~K \cite{ni1,FR}. Application of small amount of pressure or introducing chemical dopants rapidly suppresses the transition in CaFe$_{2}$As$_{2}$ and superconductivity is observed at low temperatures \cite{PC}. At higher pressures, there is a transition to the non-magnetic collapsed tetragonal ($cT$) phase \cite{newPC}, where significant reduction of the $c$-axis length is observed (see Fig.~\ref{1}). CaFe$_{2}$As$_{2}$ is also very sensitive to nonhydrostacity and the presence of superconducting state strongly depends on the way the pressure is transmitted to the sample \cite{ran,yu,AS}. Therefore the interplay of the Fermi-surface nesting and strong spin-lattice coupling is a key parameter for understanding the origin of magnetism and superconductivity in CaFe$_{2}$As$_{2}$ and other Fe-based materials in general.

The band structure calculations suggest that CaFe$_{2}$As$_{2}$ undergoes a large Fermi-surface reconstruction at the SDW/structural transitions. In the high temperature $T$ phase, the Fermi surface has strongly 2D character with two wrapped electron cylinders at the corners of the Brillouin zone (M) and two hole cylinders at the zone center ($\Gamma$) \cite{s1,s2,s3}. During transformation to the magnetic $O$ phase, the electron pocket remains quasi 2D although the hole cylinder forms a 3D oval \cite{lon,t2}. Recent angle-resolved photoemission spectroscopy (ARPES) and quantum oscillation measurements indeed show a dramatic change of the Fermi surface topology in CaFe$_{2}$As$_{2}$ at 170~K \cite{A1,A2,A3,13,14}. In the $cT$ phase, the theoretical calculations also predict a major topological change of the Fermi surface: the two electron pockets at the M point transform into one cylinder, whereas the hole pockets at $\Gamma$ disappear \cite{ty,t1,t2,t3}. In these contexts, it is important to understand the mechanism of the Fermi surface renormalization that influences the electronic properties of the competing phases in CaFe$_{2}$As$_{2}$ and its relationship to magnetism and superconductivity, especially the $cT$ phase. The nature of the $cT$ phase, however, is still not well accounted and understood and ARPES results are missing.

Here we use extensive transport, angle-resolved photoemission, and x-ray diffraction measurements to study the Fermi-surface reconstruction and complex phase equilibria in CaFe$_{2}$As$_{2}$ crystals at different crystallographic structures. By proper annealing procedure \cite{ran}, we are able to synthesize and study samples of CaFe$_{2}$As$_{2}$ that can be placed at specific parts of the phase diagram (see red solid arrows in Fig.~\ref{1}). Therefore, we can access directly the $cT$ phase and also track changes in the electronic properties similar to CaFe$_{2}$As$_{2}$ under pressure. We show that the Fermi surface of CaFe$_{2}$As$_{2}$ experiences a large reconstruction at the SDW transition, especially at the hole part of the respective Fermi surfaces. The change in the Fermi surface topology has a different character when the sample transforms to non-magnetic $cT$ phase, where we do not observe any significant contribution from the hole sheets of the Fermi surface. The pressure effect can be linked to the renormalization of the band structure governed by the strength of interband interactions, as identified by direct measurement of the Fermi surface renormalization with ARPES, and modelling.

Crystals of CaFe$_{2}$As$_{2}$ were grown out of FeAs flux with the typical size of about 2$\times$2$\times$0.1 mm$^{3}$. By proper thermal annealing CaFe$_{2}$As$_{2}$ can be tuned to a low-temperature antiferromagnetic $O$ or a non-magnetic $cT$ state, all at ambient pressure, although a bulk superconducting state in never observed \cite{ran}. We study three single crystals of CaFe$_{2}$As$_{2}$ extracted from the same batch. The first sample (s\#1) has been annealed at 350$^{o}$C for 5 days and shows antiferromagnetic ordering below $T_{N}~\simeq$~168~K. The second sample (s\#2) was annealed at 700$^{o}$C for 1 day and exhibits $T_{N}~\simeq$~116~K. The third sample (s\#3) is as-grown CaFe$_{2}$As$_{2}$ crystal showing transition to non-magnetic $cT$-phase at $T_{tr}~\simeq$~95~K (see Fig.~\ref{1}). The transport properties were measured using a Quantum Design PPMS-14 setup. The low-temperature diffraction experiments were carried out using a PANalytical X'Pert PRO MPD $x$-ray diffractometer with Cu-K$_{\alpha 1}$ radiation. ARPES measurements were performed at SLS on the SIS-X09LA beam-line. We used 82~eV photons, with instrumental resolution of $\sim$15 meV and base vacuum better than 6$\times$10$^{-11}$ mbar.

\begin{figure}[b!]
\begin{centering}
\includegraphics[width=0.35\textwidth]{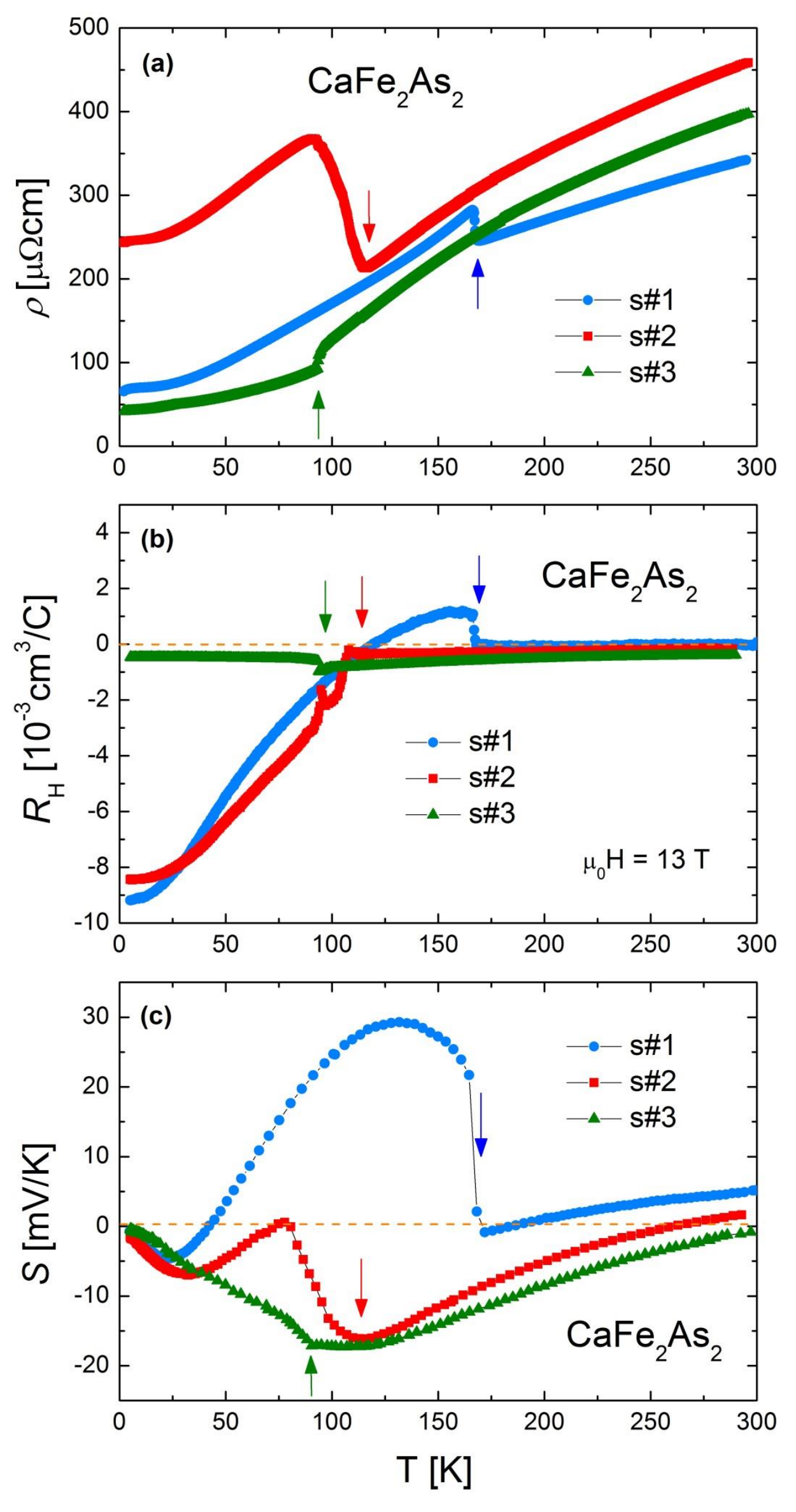}
\caption{(Color online) The temperature dependencies of the electrical resistivity (a), Hall effect (b) and Seebeck coefficient (c) of CaFe$_{2}$As$_{2}$ single crystals. All the curves were measured on cooling the samples. The arrows mark the antiferromagnetic (s\#1 and s\#2) and collapse tetragonal (s\#3) phase transitions (see text).}\label{2}
\end{centering}
\end{figure}

\begin{figure}[t!]
\begin{centering}
\includegraphics[width=0.37\textwidth]{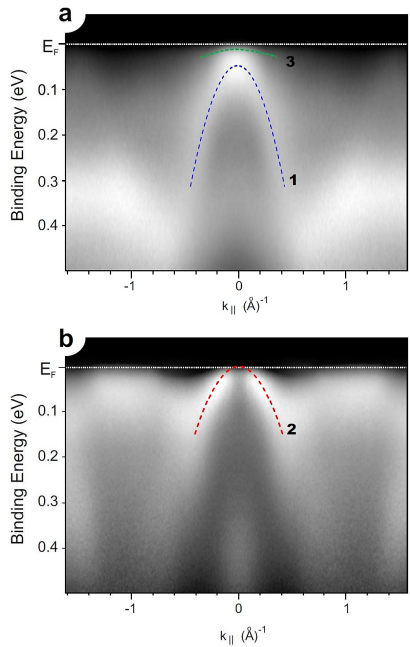}
\caption{(Color online) Band renormalization upon the $T/cT$ transition measured by ARPES. Panels a) and b) show the $\Gamma$-X direction in the Brillouin zone mapped for the $cT$ and $T$ samples, respectively. A clear shift in binding energy of the main hole-like band in the zone center may be noted. See text and \cite{Supplement} for more details.}\label{ARPES}
\end{centering}
\end{figure}

The electronic transport, especially Hall and Seebeck effects, is an effective probe of the Fermi surface volume and density of states in the vicinity of the Fermi level. Recently, the transport measurements have been used successfully to reveal details of the Fermi surface topology in heavy fermions, cuprates, and Fe-based superconductors \cite{fs1,fs2,fs3,fs4,fs5,fs6}. Figure \ref{2} shows the temperature dependence of the electrical resistivity [$\rho(T)$], Hall [$R_{H}(T)$], and Seebeck [$S(T)$] coefficients of CaFe$_{2}$As$_{2}$ crystals. It is worth noting that all the transport properties obtained for s\#1 show similar behaviors to the ones reported for Sn-flux grown CaFe$_{2}$As$_{2}$ with $T_{N}~\simeq$~170~K (see Refs.~\onlinecite{MM,FR}). The increase of $\rho(T)$ below $T_{N}$~=~168~K is consistent with the presence of SDW state in CaFe$_{2}$As$_{2}$ and can be understood by the Suezaki and Mori model \cite{sm}. At the transition temperature, both $R_{H}$ and $S$ change sign and become positive. In addition, the discontinuous nature of the transition and the presence of narrow hysteresis of the order of 2~K (see Supplemental Material \cite{Supplement} for more details) are consistent with the first-order character of the phase transition in this sample. The behavior of the Hall and Seebeck effects is characteristic of metals with electronic transport governed by electron and hole bands, in agreement with the electronic structure calculated for $T$ phase of CaFe$_{2}$As$_{2}$ \cite{ty,t1,t2,t3}. Below the transition the electronic structure is strongly reconstructed and shows a significant contribution from the hole-type of the Fermi sheets. At low temperatures, the Hall and Seebeck coefficient change sign to negative, most probably due to higher mobility of the electron carriers. A single band model provides an estimate for the concentration of free electrons $n_{H}$ of 1.2~$\times$~10$^{23}$cm$^{-3}$ and 6.8~$\times$~10$^{20}$cm$^{-3}$, at 250 and 5~K, respectively. This should be considered as the upper limit of the actual carrier concentrations. However, besides this crude approximation, the drop of $n_{H}$ below $T_{N}$ is consistent with the Fermi surface gapping scenario. All these results support recent band structure calculations \cite{ty,t2} and are consistent with ARPES and quantum oscillation measurements of s\#1, which is typically studied phase \cite{A1,A2,A3,13,14}.

The overall transport properties of s\#2 ($T_{N}~\simeq$~116~K) are similar to those of s\#1 (see Fig.~\ref{2}). The electrical resistivity shows a pronounced maximum below the magnetic and structural transition. Similarly, as shown by Hall and Seebeck measurements, below the transition $R_{H}$ and $S$ start to rise but their overall change is not as rapid as observed in s\#1 \cite{Supplement}. In addition, the width of the transitions is much wider as well as a range of the hysteretic behavior. This could point to similar but significantly weaker change of the electronic structure in this material at the transition. We attribute this behavior to the large range of the phase coexistence in this sample \cite{Supplement}, which creates a significant difficulty in studying the effect of the Fermi surface reconstruction of the hole pockets at the transition region, as it was possible in s\#1. At low temperatures, however, the Hall and Seebeck data agree well, indicating similar physics involved.

A remarkably different situation is observed for CaFe$_{2}$As$_{2}$ s\#3 phase. As shown in Fig.~\ref{2}, all the transport coefficients experience a rapid change at the transition to the $cT$ phase, accompanied by a small thermal hysteresis, both being consistent with the first order nature of the transition \cite{Supplement}. This change, however, has different character than previously observed in the s\#1 and s\#2 samples. In the tetragonal phase, both $R_{H}$ and $S$ show a similar temperature dependence to the previously discussed samples. At the $cT$ transition the electrical resistivity drops, most probably due to increase of the carrier concentration (see below). In contrast to the situation described for s\#1 and s\#2, for s\#3 and below $T_{tr}$, the Hall effect and thermoelectric power do not change sign but instead their magnitude decrease. This indicates that the Fermi surface reconstruction to the $cT$ phase does not create a significant, if any, hole-like topology as has been observed for magnetically ordered CaFe$_{2}$As$_{2}$. Furthermore, the transition to the $cT$ phase increases the electron-like Fermi surface volume in this material as shown by the Hall effect. The electron carrier concentration increases from 6.8~$\times$~10$^{21}$cm$^{-3}$ just above the transition to 1.1~$\times$~10$^{22}$cm$^{-3}$ below $T_{tr}$. It has been predicted that the electronic structure of the $T$ and $cT$ phases are surprisingly similar despite the large $c$-axis reduction \cite{ty}. Furthermore, it was concluded that the $cT$ phase has a lower energy compared to the tetragonal phase so most of the bands are just shifted to lower energy in the $cT$ phase. The results of angle-resolved photoemision study, shown in Fig.~\ref{ARPES} clearly show that indeed a band shift towards higher binding energies upon the $T/cT$ transition is the dominant band structure effect upon enetering the collapsed phase. Specifically, the bands just above the Fermi level in the $T$-phase would move below $E_{F}$ in $cT$-phase, altogether leading to disappearance of the hole pocket at the center of Brillouin zone \cite{t3}. Moreover, since only one electron-like band crosses the Fermi level along $\Gamma$-M direction, the total density of states at $E_{F}$ is reduced as compared to the $T$-phase \cite{ty,t1,t3}. Besides the band shift, we also observe change in effective mass, where effective masses for bands (1),  (2) and (3) in Fig.~\ref{ARPES} are estimated as -2.8$m_e$, -4.1$m_e$ and approximately -15-20$m_e$, respectively. This would correspond to a mass enhancement factor of about 4-5 upon transition from $T$ to $cT$ phase, assuming the band closest to the Fermi level for mass renormalization considerations. It should be noted here that while bands (1) and (2) are relatively well pronounced, band (3) shows much less spectral intensity. Also, the structure around band (2) may contain another hole-like band of similar effective mass, which was not resolved here. These points do not change our central conclusions of (i) band shift and (ii) mass enhancement upon transition.

\begin{figure}[t!]
\begin{centering}
\includegraphics[width=0.4\textwidth]{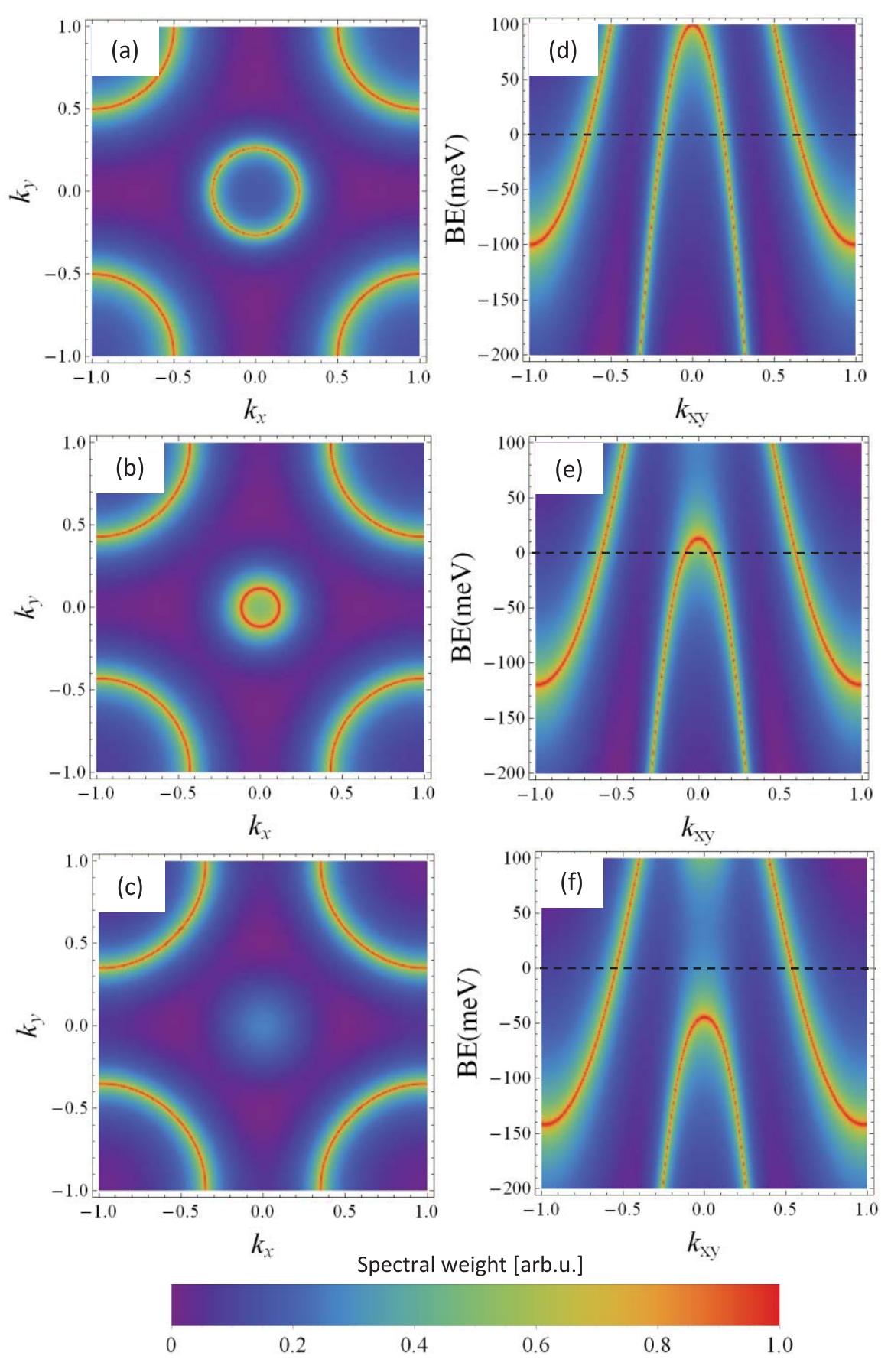}
\caption{(Color online) Modeling the band renormalization sequence. Panels  (a - c) show the $k_x - k_y$ cuts of the Fermi surface, while panels  (d - f) show the corresponding $k_{xy} - \omega $ cuts of the near- Fermi level region of the electronic band structure ($k$ -  momentum in reciprocal space). Interband  coupling strength is increased from $0$ in panels (a) and (d) to $maximum$ value in panels (c) and (f). See text and \cite{Supplement} for more details. }\label{bands}
\end{centering}
\end{figure}

The disappearance of the hole pocket at the $\Gamma$ point has a dramatic consequence for magnetism and superconductivity since interplay of the Fermi-surface nesting and strong spin-lattice coupling has been proposed to control those phenomena in Fe-based superconductors \cite{s1,s2,s3}. ARPES measurements confirms the transport results and show a distinct evolution of the band structure renormalization at the $T$ to $cT$ transition in sample s\#3, observed from the Z point in the zone center towards the $\Gamma$ point belonging to the next BCT zone. Here, a shift of the band structure towards higher binding energies is observed as a dominant effect at the transition. The observed shift is momentum dependent, spanning a range from about 50~meV for the hole-like band in the zone center, to about 100~meV for the electron-like band at the $\Gamma$ point. Shift of the hole-like band below the Fermi level is accompanied by (i) destruction of the centrally located small hole-like Fermi surface, and (ii) increase an effective mass, as indicated by the flat-like parts of renormalized spectral weight appearing at the top of the band.

We propose that the above discussed properties can be linked to the renormalization of the electronic structure driven by the evolution of electronic interband coupling constants. We approach this model from the point of view of the rigid, but momentum-dependent band shift that can be visualized by application of a single parameter, namely the interband scattering strength or coupling constant $\lambda$ \cite{model,Ortenzi}. The results of this simulation, are in very good agreement with experimental momentum resolved electronic structure renormalization through the transition obtained from ARPES, and are shown in Fig.~\ref{bands}. Here we start with a high-temperature band structure, simplified after \cite{bare_bands1, bare_bands2}, shown in panels (a) and (d). This structure is used in the model as the bare-band or non-renormalized structure, with $\lambda=0$. Upon increasing pressure or equivalently the interband coupling strength $\lambda$, we observe the central hole-like band to be shifted below the Fermi level, while its effective mass increases by about a factor of 2, in agreement with transport measurements of $T$ and $cT$ phases, and ARPES results. This observation provides a useful link between the transport and electronic properties, and is verified by photoemission experiments.

In summary, we have investigated the electronic properties of CaFe$_{2}$As$_{2}$ crystals with different magnetic/crystallographic structures. We show that the Fermi surface topology, especially the hole-part of the Fermi surface, experiences a large reconstruction at the SDW/structural transition. For the crystal that shows the transition to the non-magnetic $cT$ phase, the reconstruction of the Fermi surface is different and we do not observe significant contribution from the hole part of the Fermi surface. Furthermore, our data, especially the combination of ARPES measurements and interband scattering model, together with the evolution of the Hall and Seebeck effect and magnitude of the cotangent of Hall angle (see \cite{Supplement}), suggest that the effect of pressure on change in the Fermi surface topology results mainly in rigid-band shift to lower energy. These new results extend our understanding of this system beyond the recent band structure calculations \cite{ty,t1,t3} and form the basis of our proposed renormalized band structure model. The disappearance of the hole pockets in $cT$ phase prevents Fermi-surface nesting, which has been proposed to be necessary ingredient for magnetism and spin-fluctuation mediated superconductivity in iron-base materials \cite{s1,s2,s3}. This fully explains lack of magnetism and bulk superconductivity in collapsed tetragonal phase of CaFe$_{2}$As$_{2}$.

\begin{acknowledgments}

This work was supported by the U.S. Department of Energy, Basic Energy Sciences, Materials Sciences and Engineering Division, and in parts by the LANL LDRD program.

\end{acknowledgments}

\end{document}